\documentclass[asos,preprint]{imsart}
\usepackage{amstext,amsmath,amssymb}
\usepackage{graphicx}
\usepackage{algorithm}
\usepackage{algpseudocode}
\makeatletter
\algrenewcommand\ALG@beginalgorithmic{\footnotesize}
\makeatother
\usepackage{fmtcount}
\usepackage{float}
\usepackage{afterpage}
\usepackage[pagewise]{lineno}
\usepackage[dvipsnames,usenames]{color}
\usepackage[normalem]{ulem}
\RequirePackage[authoryear]{natbib}
\usepackage{tabularx}
\usepackage{rotating}
\usepackage{enumerate}
\usepackage{graphicx}

\usepackage{caption}
\usepackage{subcaption}
\usepackage[maxfloats=40]{morefloats}

\usepackage[resetlabels,labeled]{multibib}
\newcites{sec}{Referenceswss}

\newcommand{\norm}[1]{\left|\left|#1\right|\right|}
\newcommand{\abs}[1]{\left|#1\right|}
\DeclareMathOperator*{\trace}{trace}
\DeclareMathOperator*{\sign}{sign}
\usepackage{tkz-graph}
\usetikzlibrary{arrows,positioning,automata}

\newtheorem{lemma}{Lemma}

\numberwithin{equation}{section}
\numberwithin{lemma}{section}
\numberwithin{thm}{section}

\begin{document}

\begin{frontmatter}

\title{Estimation of Gaussian directed acyclic graphs using partial ordering information with an application to dairy cattle data}
\runtitle{Estimation of partially ordered DAGs}
\author{\fnms{Syed} \snm{Rahman}\ead[label=e1]{shr264@ufl.edu}},
\author{\fnms{Kshitij} \snm{Khare} \ead[label=e2]{kdkhare@stat.ufl.edu}},
\author{\fnms{George} \snm{Michailidis} \ead[label=e3]{gmichail@ufl.edu}},
\author{\fnms{Carlos} \snm{Martinez} \ead[label=e4]{carlosmn@ufl.edu}},
\and
\author{\fnms{Juan} \snm{Carulla} \ead[label=e5]{jecarullaf@unal.edu.co}}

\runauthor{S. Rahman et al.}

\affiliation{University of Florida, Corporacian Colombiana de Investigacian Agropecuaria, and Universidad Nacional de 
Colombia}

\begin{abstract}
Estimating a directed acyclic graph (DAG) from observational data represents a canonical learning problem and has 
generated a lot of interest in recent years. Research has focused mostly on the following two cases: when no 
information regarding the ordering of the nodes in the DAG is available, and when a domain-specific complete 
ordering of the nodes is available. In this paper, motivated by a recent application in dairy science, we develop a 
method for DAG estimation for the middle scenario, where partition based partial ordering of the nodes is known 
based on domain-specific knowledge. We develop an efficient algorithm that solves the posited problem, coined 
Partition-DAG. Through extensive simulations using the DREAM3 Yeast data, we illustrate that Partition-DAG 
effectively incorporates the partial ordering information to improve both speed and accuracy. We then illustrate the 
usefulness of Partition-DAG by applying it to recently collected dairy cattle data, and inferring relationships between 
various variables involved in dairy agroecosystems. 
\end{abstract}


\end{frontmatter}

\section{Introduction} \label{introduction}


The problem of estimating a directed acyclic graph (DAG) from high-dimensional observational data has attracted a lot of attention recently in the statistics
and machine learning literature, due to its importance in a number of application areas including molecular biology. In the latter area, high throughput
techniques have enabled biomedical researchers to profile biomolecular data to better understand causal molecular mechanisms involved in gene regulation and protein signaling (\cite{emmert2014gene}). Further, it provided the impetus for the develoment of numerous approaches for tackling the problem 
-  see for example the review paper \cite{marbach2012wisdom} and references therein.

This is a challenging learning problem in its general form. It stems from the fact that in order to reconstruct a DAG from data, one has to consider
all possible orderings of nodes and score the resulting network structures based on evidence gleaned from the available data. The computational complexity of obtaining all possible orderings of a set of nodes in a directed graph is exponential in the size of the graph. In certain applications, one may have
access to a complete topological ordering, which renders the problem computationally tractable as discussed below. An interesting question arises
of what advantages, the availability of external information on partial orderings of nodes, brings to solving the problem. This is the key issue addressed in
this study, motivated by an application on dairy operations, where such information can be reliably obtained from operators as explained next.

Dairy cattle operations are characterized by complex interactions between 
several factors which determine the success of these systems. Most of these operationsresult in the collection of large amounts of data
that are usually analyzed using univariate statistical models for certain variables of interest; therefore, information from relationships between 
these variables is ignored. In addition, due to the structure of a dairy cattle agroecosystem, it is of great 
interest to carry out data analysis that permits to implement a systemic approach (\cite{Jalvingh:1992, Thornley:France:2007}).
Moreover, due to their high relevance when making management decisions and recommendations, knowing not only 
interaction patterns, but also causal relationships between the components of dairy production systems, is a problem of 
current interest, and has the potential to have a marked impact on the dairy industry. In any of these systems (such as the data 
analyzed in Section \ref{dairydataunoc}), causal relationships between selected pairs of variables  are reliably known. The 
statistical task then, is to leverage these known relationships and the observed data to estimate the underlying 
network of relationships between the variables under consideration. 

In particular, suppose ${\bf Y}_1,...,{\bf Y}_n \in \mathbb{R}^p$ are i.i.d. random vectors from a multivariate distribution with 
mean ${\bf 0}$ and covariance matrix $\Sigma = \Omega^{-1}$. In sample deprived settings, an effective and popular method 
for estimating $\Sigma$ imposes sparsity on the entries of $\Sigma$ (covariance graph models), or $\Omega$ (graphical 
models), or appropriate Cholesky factors of $\Omega$ (directed acyclic graph models or Bayesian networks). The choice of 
an appropriate model often depends on the application. 

In this study, our focus is on learning DAG from high-dimensional data, assuming {\em sparsity} in
an appropriate Cholesky factor of $\Omega$. In particular, consider the 
factorization of the inverse covariance matrix \(\Omega = B^t B\), where $B$ can be converted to a lower triangular matrix 
with positive diagonal entries through a permutation of rows and columns. The sparsity pattern of \(B\) 
gives us a \textit{directed acyclic graph (DAG)}, which is defined as \(G = (V,E)\)
where \(V = \{1,...,p\}\) and \(E = \{i \to j: B_{ij} \neq 0 \}\). The assumption $B_{ij} = 0$ corresponds to 
assuming an appropriate conditional independence under Gaussianity, and corresponds to assuming 
an appropriate conditional correlation being zero in the general setting. 

There are two main lines of work that have dealt with DAG estimation in the Gaussian 
framework - one where the permutation that makes B lower triangular is known and one where it is 
unknown. We brifely discuss them below. As previously mentioned, nn many applications, a natural ordering, such as time based 
or location based ordering, of the variables presents itself, and hence a natural choice for the 
permutation which makes $B$ lower triangular is available. Penalized likelihood methods, which use 
versions of the $\ell_1$ penalty, and minimize the respective objective functions over the space of 
lower triangular matrices, have been developed in \cite{HLPL:2006, Shajoie:Michalidis:2010, KORR:2017}. 
Bayesian methods for this setting have been developed in 
\cite{CKG:2017, Altamore:Consonni:2013, Consonni:2017}. For many of these 
methods, high dimensional consistency results for the model parameters have also been established. 

If the permutation/ordering that makes $B$ lower triangular is unknown, then the problem becomes 
significantly more challenging, both computationally and theortically. In this setting, several score-
based, constraint-based and hybrid algorithms for estimating the underlying CPDAG~\footnote{the 
permutation is not identifiable from observational data, but one can recover an equivalence class of 
DAGs, refered to as completed partially DAG or CPDAG, from observational data.} have been 
developed and studied in the literature (\cite{PCR:2001, GH:2013, LB:1994, H:1995, C:2003, EW:2008, Z:2011, KB:2007, T:2006, G:2011, G:2012, vandeGeer:Buhlmann:2013}). See \cite{Aragam:Zhou:2015} for an excellent and detailed review. 
Recently, \cite{Aragam:Zhou:2015} have developed a penalized likelihood approach called CCDr for sparse estimation of 
$B$, which has been shown to be significantly more computationally scalable than previous approaches. 

However, in some applications, such as the dairy cattle data studied in Section \ref{dairydataunoc}, domain-specific 
information regarding the variables in available, which allows for a partition of the variables into 
sets $V_1, V_2, \cdots, V_k$ such that any possible edge from a vertex in $V_i$ to a vertex in $V_j$ 
is directed from $v_i$ to $v_j$ if $i < j$. However, the ordering of the variables in the same subset is 
not known, and has to be inferred from the data. This setting is not subsumed in the previous 
approaches mentioned above, and falls somewhere in the middle of the two extremes of having 
complete information regarding the ordering, and having no information regarding the ordering. The 
goal of this paper is to develop a hybrid approach for DAG estimation which leverages the partition 
based partial ordering information. We will also show that using the partition information leads to a reduction 
in the number of computations, and more importantly allows for parallel processing, unlike CCDr. 
This can lead to significant improvement in computational speed and statistical performance. 

The remainder of the paper is organized as follows. In Section \ref{method}, we develop, and describe in detail, our 
hybrid algorithm called {\it Partition-DAG}. In Section \ref{simulations}, we perform a detailed experimental study to 
evaluate and understand the performance of the proposed algorithm. In Sections \ref{partial:ordering:no:ordering}, 
\ref{fine:partition:coarse:partition} and \ref{informative:ordering:noninformative:ordering}, we use known DAGs from 
the DREAM3 competition (\cite{dream1, dream2, dream3}) and perform an extensive simulation study to explore the 
effectiveness/ability of Partition-DAG to incorporate the ordering information, and how this ability changes with more/less 
informative partitions. In Section \ref{randomdags}, we perform a similar simulation study, this time using randomly 
generated DAGs with more number of variables. Finally, in Section \ref{dairydataunoc} we analyze dairy cattle data 
recently gathered by Universidad Nacional de Colombia using the proposed Partition-DAG approach.

\section{DAG estimation using partition information and the corresponding Partition-DAG algorithm}\label{method}

\noindent
In order to understand how one can leverage the partial ordering information, it is crucial to understand the workings, 
similarities, and differences of the Concave penalized Coordinate Descent with reparameterization (CCDr) 
\cite{Aragam:Zhou:2015} and the Convex Sparse Cholesky Selection (CSCS) \cite{KORR:2017} algorithms, 
which are state of the art in terms on computational scalability and convergence for the boundary settings with 
completely unknown and completely known variable ordering respectively. 

The CSCS algorithm is derived under the setting where a domain-specific ordering of the variables which makes 
$B$ lower triangular is known. Hence, the DAG estimation problem boils down to estimating the sparsity pattern in $L$, the 
lower triangular permuted version of $B$. In other words, $L$ is a lower triangular matrix with positive diagonal entries such that 
$\Omega = L^t L$. The objective function for CSCS is
\[
\begin{aligned}
Q_{cscs}(L) &= \trace(\Omega S) - \frac{1}{2} \log \abs{\Omega} + \lambda \sum_{1 \leq j < i \leq p} \abs{L_{ij}} \\
&= \trace(L^t L S) -  \log \abs{L} + \lambda \sum_{1 \leq j < i \leq p} \abs{L_{ij}}, 
\end{aligned}
\]

\noindent
where $S$ is the sample covariance matrix. The first two terms in $Q_{cscs}$ correspond to the Gaussian log-likelihood, and the 
third term is an $\ell_1$ penalty term which induces sparsity in the lower triangular matrix $L$. 

The CCDr algorithm is derived under the setting where there is no knowledge about the permutation that makes $B$ lower 
triangular.  Here \(\Omega = B^tB\), where $B$ varies over the space of $p \times p$ matrices such that by permuting the rows and 
columns of $B$ we can reduce it to a lower triangular matrix with positive diagonal entries. More formally, denoting $S_p$ to be the 
group of permutations of $\{1,2, \cdots, p\}$, we assume that $B \in \mathcal{B}_p$, where 
$$
\mathcal{B}_p = \{B: \exists \sigma \in S_p \mbox{ such that } B_{\sigma(i) \sigma(j)} = 0 \mbox{ if } i < j\}. 
$$

\noindent
The CCDr-\(\ell_1\) method uses the following objective function: \[
\begin{aligned}
Q_{CCDr}(B) &= \trace(\Omega S) - \frac{1}{2} \log \abs{\Omega} + \lambda \sum_{1 \leq i \neq j \leq p} \abs{B_{ij}} \\
&=  \trace(B^t B S) - \frac{1}{2} \log \abs{B^t B} + \lambda \sum_{1 \leq i \neq j \leq p} \abs{B_{ij}}
\end{aligned}
\] Exactly like CSCS the first two terms correspond to the Gaussian log-likelihood, while the third term tries to impose sparsity on $B$. 

While objective functions for CSCS and CCRr-\(\ell_1\) look
identical, the algorithms to minimize the two objective functions are very different due to the fact that we are minimizing over 
different spaces. Both objective functions are convex, but CSCS has the added advantage that the range for \(Q_{cscs}\), 
which is the set of lower triangular matrices with positive diagonal entries, is convex as well. This leads to a convex problem 
(though not strictly convex for $n < p$) and we can establish that the sequence of iterates converges to a global minimum of 
the objective function. However, the range for \(Q_{CCDr}(B)\), which is the set of matrices that can be converted to a lower triangular matrix with positive diagonal entries through a permutation of the rows and columns, is not convex and general results in the literature (at best) only guarantee convergence of the sequence of iterates to a local minimum of the objective function. In addition, while CSCS can be broken down into \(p\) parallelizable problems
the same can not be said for CCDr, which leads to a significant computational disadvantage for CCDr. 
Finally, asymptotic consistency for the general setting (with no
restrictions on conditional variances) for CCDr isn't available as yet, whereas Theorem 4.1 in \cite{KORR:2017} establishes 
both model selection and estimation consistency for CSCS. See \cite{KORR:2017} for a more detailed comparison between 
these two algorithms.

\noindent
As stated in the introduction, in many applications, additional data can give information about partitions of the variables where 
we have prior knowledge about the direction of the edges between partitions, but not within partitions (for example, the 
dairy cattle data in Section \ref{dairydataunoc}, or gene knock-out data or more general perturbations data 
\cite{SJKM:2014}). We will now discuss how one can create a hybrid algorithm from CSCS and CCDr where we 
incorporate this information for DAG estimation. 

\subsection{The case with two partition blocks} \label{sec:two:blocks}

\noindent
For simplicity we will initially work with the case where the variables, 
\(V =\{1,...,p\}\), are divided into two groups \(V_1 = \{1,...,m\}\) and \(V_2 = \{m+1,...,p\}\) such that we cannot have an edge 
from a node in \(V_2\) to one in \(V_1\), but can have one from a node in \(V_1\) to one in \(V_2\). Hence, \(\forall j \in V_1, \forall k \in 
V_2\) we have that \( B_{kj} = 0\). This implies that B has the block triangular form 
\begin{align} \label{eq:b}
B = \begin{pmatrix} \mathbf{B}_{11} &  \mathbf{0}\\
\mathbf{B}_{21} & \mathbf{B}_{22} \\ 
\end{pmatrix}
\end{align} 

\noindent
The diagonal blocks ${\bf B}_{11}$ and ${\bf B}_{22}$ are constrained so that each matrix is a permuted version of lower triangular 
matrices, i.e., ${\bf B}_{11} \in \mathcal{B}_m$ and ${\bf B}_{22} \in \mathcal{B}_{p-m}$, the entries of the off-diagonal block 
${\bf B}_{12}$ are all zero. However, there are no constraints on the 
off-diagonal block ${\bf B}_{21}$. Similar to CCDr, we consider a Gaussian log-likelihood based objective function, denoted 
by $Q_{PDAG}$, given by 
$$
Q_{PDAG} ({\bf B}) = \trace(\Omega S) - \frac{1}{2} \log \abs{\Omega} + \lambda \sum_{1 \leq i \neq j \leq p} \abs{B_{ij}}. 
$$

\noindent
Here, our goal is to minimize the above function over the space $\widetilde{\mathcal{B}}_p$, defined by 
$$
\widetilde{\mathcal{B}}_p = \{{\bf B}: {\bf B}_{11} \in \mathcal{B}_m, {\bf B}_{22} \in \mathcal{B}_{p-m}, {\bf B}_{12} = {\bf 0}\}, 
$$

\noindent
as opposed to CCDr, where the goal is to minimize over the space $\mathcal{B}_p$. Note that since $\mathcal{B}_m$ and 
$\mathcal{B}_{p-m}$ are not convex sets, $\widetilde{\mathcal{B}}_p$ is also not a convex set. 

\subsubsection{A roadmap for the algorithm}

\noindent
As in CCDr and CSCS, we pursue a coordinate-wise minimization approach. At each iteration, we cycle through minimizing 
$Q_{PDAG}$ with respect to each non-trivial element of $B$ (fixing the other entries at their current values). The minimizing 
value is then set as the new value of the corresponding element. We repeat the iterations until the difference between the 
$B$ values at two successive iterations falls below a user-defined threshold. 

Hence, for implementing coordinate-wise minimization, we need to understand how to minimize $Q_{PDAG}$ with respect to 
an arbitrary element $B_{ij}$ of ${\bf B}$ given all the other elements. Using straightforward calculations, we get 
\begin{align}
&Q_{PDAG}({\bf B}) \nonumber \\
&=  \trace({\bf B}^t {\bf B} S) - \frac{1}{2} \log \abs{{\bf B}^t {\bf B}} + \lambda \sum_{1 \leq i \neq j \leq p} \abs{B_{ij}}  \nonumber \\
\label{eq:pdag} &=  \sum_{i = 1}^{p } \Big( \big(\sum_{h=1}^p S_{hh}B_{ih}^2 + 2 \sum_{k = 1}^{p-1} \sum_{l=k+1}^p S_{kl}B_{ik}
B_{il} \big) -  \log B_{ii} + \lambda \sum_{j = 1, j \neq i}^{p} \abs{B_{ij}} \Big) 
\end{align}

\noindent
Given the nature of the constraints on each block of ${\bf B} \in \widetilde{\mathcal{B}}_p$, we consider three different cases. 
\begin{itemize}
\item {\bf Case I: (Diagonal entries - $B_{ii}$)} It follows from (\ref{eq:pdag}) that $Q_{PDAG}$ is the sum of quadratic and logarithmic 
terms in a given diagonal entry $B_{ii}$ (treating other entries as fixed). In particular, 
$$
Q_{PDAG} (B_{ii}) = S_{ii} B_{ii}^2 + 2 B_{ii} \sum_{k = 1, k \neq i }^p S_{ik} B_{ik} - \log B_{ii} + \mbox{ terms independent 
of } B_{ii}. 
$$

\noindent
Simple calculus shows that the unique minimizer (with respect to $B_{ii}$) of the above function is given by 
\begin{align} \label{eq:bii}
\hat{B}_{ii} = \frac{-2\sum_{k = 1, k \neq i }^{p} S_{ik} B_{ik} + \sqrt{4(\sum_{k = 1, k \neq i }^{p} S_{ik} B_{ik} )^2 + 8 S_{ii} } }{2 
S_{ii} }
\end{align}

\item {\bf Case II: (Off-diagonal entries in ${\bf B}_{21}$, the CSCS case)} Consider $B_{ij}$, where $m+1 \leq i \leq p$ and $1 \leq j \leq m$. Since 
$B_{ji} = 0$, it follows from (\ref{eq:pdag}) that $Q_{PDAG}$ is the sum of quadratic and absolute value terms in $B_{ij}$ 
(treating other entries as fixed). In particular, 
\begin{align} \label{eq:offdiag}
Q_{PDAG} (B_{ij}) = S_{jj} B_{ij}^2 + 2 B_{ij} \sum_{k = 1, k \neq j}^{p} S_{jk} B_{ik} + \lambda \abs{B_{ij}} + \mbox{ terms 
independent of } B_{ij}. 
\end{align}

\noindent
Simple calculus shows that the unique minimizer (with respect to $B_{ij}$) of the above function is given by 
\begin{align} \label{eq:bik}
\hat{B}_{ij} = \mathcal{S} \Big(\frac{-\sum_{k = 1, k \neq j}^{p} S_{jk} B_{ik}}{2S_{jj}}, \frac{\lambda}{4S_{jj}}\Big),
\end{align}

\noindent
where \(\mathcal{S}(x,\lambda) = \sign(x) \max\{\abs{x}-\lambda,0\}\). This step exactly resembles Case a typical step of the 
CSCS algorithm. 

\item {\bf Case III: (Off-diagonal entries in ${\bf B}_{11}$ and ${\bf B}_{22}$, the CCDr case)} Consider $B_{ij}$, where $1 
\leq i \neq j \leq m$ or $m+1 \leq i \neq j \leq p$. Since ${\bf B}_{11} \in \mathcal{B}_m$ and ${\bf B}_{22} \in 
\mathcal{B}_{p-m}$, it follows that at most one of $B_{ij}$ or $B_{ji}$ is non-zero. So as in CCDr \cite{Aragam:Zhou:2015}, 
we will jointly minimize $Q_{PDAG}$ as a function of $(B_{ij}, B_{ji})$. This can be done as follows. If adding a non-zero 
value for $B_{ij}$ violates the DAG constraint, or equivalently the constraint that ${\bf B}_{11} \in \mathcal{B}_m$ and 
${\bf B}_{22} \in \mathcal{B}_{p-m}$, then we set $B_{ij} = 0$, and then minimize $Q_{PDAG}$ as a function of $B_{ji}$ 
and update the $B_{ji}$ entry as specified in (\ref{eq:offdiag}), (\ref{eq:bik}) with the roles of $i$ and $j$ exchanged). If 
adding a non-zero value for $B_{ji}$ violates the DAG constraint, then we set $B_{ji} = 0$, and then minimize $Q_{PDAG}$ 
as a function of $B_{ij}$ and update the $B_{ij}$ entry as specified in (\ref{eq:offdiag}), (\ref{eq:bik}). However, it is possible 
that neither \(\abs{B_{ij}} > 0\) or \(\abs{B_{ji}} > 0\) violates the DAG constraint. In that case, we compute $\hat{B}_{ij}$ and 
$\hat{B}_{ji}$ using appropriate versions of (\ref{eq:bik}), pick the one that makes a larger contribution towards minimizing 
$Q_{PDAG}$, and set the other one to zero. This step exactly resembles a typical step of the CCDr-$\ell_1$ algorithm. 
\end{itemize}

\noindent
The resulting coordinatewise minimization algorithm for $Q_{PDAG}$, called Partition-DAG, which repeatedly iterates through 
all the entries of ${\bf B}$ based on the three cases discussed above, is provided in Algorithm \ref{alg:pdag}. 
{\it Case II and Case III, which correspond to typical steps of the CSCS and CCDr algorithm respectively, demonstrate why  
we regard the Partition-DAG algorithm as a hybrid of these two algorithms.}
\afterpage{%
\thispagestyle{empty}
   \scalebox{0.9}{%
    \begin{minipage}{\textwidth}
\begin{algorithm}[H]
\begin{algorithmic}
\State Set $B^{o} = I_p$
\State Set $\epsilon > 0$
\Comment{Optimizing $B_{11}$}
\While {$\norm{B^{n} - B^{o}}_{\infty} \geq \epsilon$}
\State Set $B^{o} = B^{n}$
\For{i = 1,...,m}
\State 
$$
\hat{B}_{ii}^{n} = \frac{-2\sum_{k = 1, k \neq i }^{m} S_{ik} B_{ik}^{n} + \sqrt{4(\sum_{k = 1, k \neq i }^{m} S_{ik} B_{ik}^{n} )^2 + 8 S_{ii} } }{2 
S_{ii} }
$$
\EndFor
\For{i = 1,...,m}
\For{j = 1,...,m}
\State Check if adding edge $j \to i$ or $i \to j$ violates DAG.
\State If $i \to j$ violates DAG, set $B_{ji}^{n} = 0$ and
$$
\hat{B}_{ij}^{n} = \mathcal{S} \Big(\frac{-\sum_{k = 1, k \neq j}^{m} S_{jk} B_{ik}^{n}}{2S_{jj}}, \frac{\lambda}{4S_{jj}}\Big)
$$
\State \State If $j \to i$ violates DAG, reverse indices and repeat the above steps.
\State Otherwise, if neither violates the DAG, set $q_1 = Q_{PDAG}(B^{n})$ evaluated with $B_{ij}^n = \hat{B}_{ij}$ and $B_{ji}^n = 0$ and  $q_2 = Q_{PDAG}(B^{n})$ evaluated with $B_{ji}^n = \hat{B}_{ji}$ and $B_{ij}^n = 0$
\State If $q_1 > q_2$, set $B_{ij}^{n} = 0$ and $B_{ji}^{n} = \hat{B}_{ji}$. Otherwise, set $B_{ji}^{n} = 0$ and calculate $B_{ij}^{n} = \hat{B}_{ij}$. 
\EndFor
\EndFor
\EndWhile
\Comment{Optimizing $B_{22}$}
\While {$\norm{B^{n} - B^{o}}_{\infty} \geq \epsilon$}
\State Set $B^{o} = B^{n}$
\For{i = m+1,...,p}
\State 
$$
\hat{B}_{ii}^{n} = \frac{-2\sum_{k = m+1, k \neq i }^{p} S_{ik} B_{ik}^{n} + \sqrt{4(\sum_{k = m+1, k \neq i }^{p} S_{ik} B_{ik}^{n} )^2 + 8 S_{ii} } }{2 
S_{ii} }
$$
\EndFor
\For{i = m+1,...,p}
\For{j = m+1,...,p}
\State Check if adding edge $j \to i$ or $i \to j$ violates DAG.
\State If $i \to j$ violates DAG, set $B_{ji}^{n} = 0$ and
$$
\hat{B}_{ij}^{n} = \mathcal{S} \Big(\frac{-\sum_{k = m+1, k \neq j}^{p} S_{jk} B_{ik}^{n}}{2S_{jj}}, \frac{\lambda}{4S_{jj}}\Big)
$$
\State \State If $j \to i$ violates DAG, reverse indices and repeat the above steps.
\State Otherwise, if neither violates the DAG, set $q_1 = Q_{PDAG}(B^{n})$ evaluated with $B_{ij}^n = \hat{B}_{ij}$ and $B_{ji}^n = 0$ and  $q_2 = Q_{PDAG}(B^{n})$ evaluated with $B_{ji}^n = \hat{B}_{ji}$ and $B_{ij}^n = 0$
\State If $q_1 > q_2$, set $B_{ij}^{n} = 0$ and $B_{ji}^{n} = \hat{B}_{ji}$. Otherwise, set $B_{ji}^{n} = 0$ and calculate $B_{ij}^{n} = \hat{B}_{ij}$.
\EndFor
\EndFor
\Comment{$B_{21}$}
\For{i = m+1,...,p}
\For{j = 1,...,m}
\State Set $B_{ji}^{n} = 0$ and
$$
\hat{B}_{ij}^{n} = S \Big(\frac{\sum_{k = 1, k \neq j}^{p} S_{jk} B_{ik}^{n}}{2S_{jj}}, \frac{\lambda}{4S_{jj}}\Big)
$$
\EndFor
\EndFor
\EndWhile
\end{algorithmic}
\caption{Partition-DAG algorithm with 2 blocks}
\label{alg:pdag}
\end{algorithm}
\end{minipage}
}
}


\subsection{The case with multiple partition blocks} \label{sec:multiple:blocks}

\noindent
Algorithm \ref{alg:pdag} can be easily generalized to the case where the variables are partitioned into $R$ blocks, say, $V_1, 
V_2, \cdots, V_R$, such that any edge from a node $u$ in $V_i$ to a node $v$ in $V_j$ is directed from $u$ to $v$, if $i < j$. 
However, the ordering within each $V_i$ is not known. In particular, let $V_i = \{m_{i-1}+1, \cdots, m_i\}$, where $m_0 = 0$ 
and $m_R = p$. Under these constraints, the matrix $B$ has a block lower triangular structure, which can be denoted as 
follows. 
\begin{align}
B = \begin{pmatrix} \mathbf{B}_{11} & \mathbf{0} & \hdots &  \mathbf{0} \\
\mathbf{B}_{21} & \mathbf{B}_{22} & \hdots &  \mathbf{0} \\ 
\vdots & \vdots & \ddots &  \vdots \\ 
\mathbf{B}_{R1} & \mathbf{B}_{R2} & \hdots &   \mathbf{B}_{RR} \\ 
\end{pmatrix}
\end{align}

\noindent
In particular, the parameter $B$ lies in the space $\widetilde{\mathcal{B}}_{p,R}$ given by 
$$
\widetilde{\mathcal{B}}_{p,R} = \{{\bf B}: {\bf B}_{ii} \in \mathcal{B}_{m_i - m_{i-1}}, \; {\bf B}_{rs} = {\bf 0} \mbox{ if } 1 \leq r < s \leq 
R\}. 
$$

\noindent
We again use a coordinate-wise minimization approach for minimizing $Q_{PDAG}$ over $\widetilde{\mathcal{B}}_{p,R}$. 
Similar to the two partition block case, the coordinate-wise minimizations can be divided into three cases. 
\begin{itemize}
\item The first case deals with a diagonal entries $B_{ii}$, and the unique minimizer has exactly the same form as in 
(\ref{eq:bii}). 
\item The second case (the CSCS case) deals with off-diagonal entries $B_{ij}$ which belong to one of the lower triangular 
blocks ${\bf B}_{rs}$ with $1 \leq s < r \leq R$, and the unique minimizer has  exactly the same form as in (\ref{eq:offdiag}). 
\item Finally, the third case (the CCDr case) deals with off-diagonal entries $B_{ij}$ which belong to one of the diagonal 
blocks ${\bf B}_{ii}$ with $1 \leq r \leq R$, and the unique minimizer has eactly the same form as in (\ref{eq:bik}). The 
algorithm, using the steps described above, is provided as Algorithm \ref{alg:pdagR} in the appendix. 
\end{itemize}

\noindent
Note that while $Q_{PDAG}$ is jointly convex in $B$, the domain of minimization $\widetilde{\mathcal{B}}_{p,R}$ 
is not a convex set. Hence, to the best of our knowledge, existing results in the literature do not imply convergence 
of the coordinate-wise minimization algorithm (Algorithm \ref{alg:pdagR}). Using standard arguments (for example, 
similar to Theorem 4.1 of \cite{Tseng:2001}), the following result can be established. 
\begin{lemma}
Assuming that all the diagonal entries of $S$ are positive, any cluster point of the sequence of iterates produced by 
Algorithm \ref{alg:pdagR} is a stationary point of $Q_{PDAG}$ in $\widetilde{\mathcal{B}}_{p,R}$. 
\end{lemma}

\afterpage{%
\thispagestyle{empty}
   \scalebox{0.9}{%
    \begin{minipage}{\textwidth}
\begin{algorithm}[H]
\begin{algorithmic}
\State Set $B^{o} = I_p$
\State Set $\epsilon > 0$
\Comment{Optimizing $B_{11}$}
\While {$\norm{B^{n} - B^{o}}_{\infty} \geq \epsilon$}
\State Set $B^{o} = B^{n}$
\For{$i = 1,...,m_1$}
\State 
$$
\hat{B}_{ii}^{n} = \frac{-2\sum_{k = 1, k \neq i }^{m_1} S_{ik} B_{ik}^{n} + \sqrt{4(\sum_{k = 1, k \neq i }^{m_1} S_{ik} B_{ik}^{n} )^2 + 8 S_{ii} } }{2 
S_{ii} }
$$
\EndFor
\For{$i = 1,...,m_1$}
\For{$j = 1,...,m_1$}
\State Check if adding edge $j \to i$ or $i \to j$ violates DAG.
\State If $i \to j$ violates DAG, set $B_{ji}^{n} = 0$ and
$$
\hat{B}_{ij}^{n} = \mathcal{S} \Big(\frac{-\sum_{k = 1, k \neq j}^{m_1} S_{jk} B_{ik}^{n}}{2S_{jj}}, \frac{\lambda}{4S_{jj}}\Big)
$$
\State \State If $j \to i$ violates DAG, reverse indices and repeat the above steps.
\State Otherwise, if neither violates the DAG, set $q_1 = Q_{PDAG}(B^{n})$ evaluated with $B_{ij}^n = \hat{B}_{ij}$ and $B_{ji}^n = 0$ and  $q_2 = Q_{PDAG}(B^{n})$ evaluated with $B_{ji}^n = \hat{B}_{ji}$ and $B_{ij}^n = 0$
\State If $q_1 > q_2$, set $B_{ij}^{n} = 0$ and $B_{ji}^{n} = \hat{B}_{ji}$. Otherwise, set $B_{ji}^{n} = 0$ and calculate $B_{ij}^{n} = \hat{B}_{ij}$. 
\EndFor
\EndFor
\EndWhile
\For{r = 2,...,R}
\Comment{Optimizing $r^{th}$ Block Row}
\While {$\norm{B^{n} - B^{o}}_{\infty} \geq \epsilon$}
\State Set $B^{o} = B^{n}$
\For{$i = m_{r-1}+1,...,m_{r}$}
\State 
$$
\hat{B}_{ii}^{n} = \frac{-2\sum_{k = m_{r-1}+1, k \neq i }^{m_{r}} S_{ik} B_{ik}^{n} + \sqrt{4(\sum_{k = m_{r-1}+1, k \neq i }^{m_{r}} S_{ik} B_{ik}^{n} )^2 + 8 S_{ii} } }{2 
S_{ii} }
$$
\EndFor
\For{$i = m_{r-1}+1,...,m_{r}$}
\For{$i = m_{r-1}+1,...,m_{r}$}
\State Check if adding edge $j \to i$ or $i \to j$ violates DAG.
\State If $i \to j$ violates DAG, set $B_{ji}^{n} = 0$ and
$$
\hat{B}_{ij}^{n} = \mathcal{S} \Big(\frac{-\sum_{k = m_{r-1}+1, k \neq j}^{m_{r}} S_{jk} B_{ik}^{n}}{2S_{jj}}, \frac{\lambda}{4S_{jj}}\Big)
$$
\State \State If $j \to i$ violates DAG, reverse indices and repeat the above steps.
\State Otherwise, if neither violates the DAG, set $q_1 = Q_{PDAG}(B^{n})$ evaluated with $B_{ij}^n = \hat{B}_{ij}$ and $B_{ji}^n = 0$ and  $q_2 = Q_{PDAG}(B^{n})$ evaluated with $B_{ji}^n = \hat{B}_{ji}$ and $B_{ij}^n = 0$
\State If $q_1 > q_2$, set $B_{ij}^{n} = 0$ and $B_{ji}^{n} = \hat{B}_{ji}$. Otherwise, set $B_{ji}^{n} = 0$ and calculate $B_{ij}^{n} = \hat{B}_{ij}$.
\EndFor
\EndFor
\For{$i = m_{r-1}+1,...,m_{r}$}
\For{$j = 1,...,m_{r-1}$}
\State Set $B_{ji}^{n} = 0$ and
$$
\hat{B}_{ij}^{n} = S \Big(\frac{\sum_{k = 1, k \neq j}^{m_{r}} S_{jk} B_{ik}^{n}}{2S_{jj}}, \frac{\lambda}{4S_{jj}}\Big)
$$
\EndFor
\EndFor
\EndWhile
\EndFor
\end{algorithmic}
\caption{Partition-DAG algorithm with R blocks}
\label{alg:pdagR}
\end{algorithm}
\end{minipage}
}
}

\subsection{Advantages of Partition-DAG} \label{partition:advantages}

\noindent
We now discuss some of the computational and statistical advantages of using the partition based ordering information in the 
DAG estimation algorithms derived in this paper. 
\begin{enumerate}
\item (Parallelizability) Consider the general multiple block case described in Section \ref{sec:multiple:blocks}. After some 
manipulations, it can be shown that the objective function $Q_{PDAG}$ can be decomposed as a sum of $R$ functions, where 
each function exclusively uses entries  from a distinct block row of $B$. In particular, it can be shown that 
$$
Q_{PDAG} ({\bf B}) = \sum_{r=1}^R Q_r ({\bf B}_{r1}, {\bf B}_{r2}, \cdots, {\bf B}_{rr}), 
$$

\noindent
where 
\begin{align*}
Q_r ({\bf B}_{r1}, {\bf B}_{r2}, \cdots, {\bf B}_{rr}) &= \sum_{i = m_{r-1}+1}^{m_r} \Big( \big(\sum_{h=1}^{m_r} S_{hh} B_{ih}^2 + 2 \sum_{k = 1}^{m_r-1} \sum_{l=k+1}^{m_r} S_{kl}B_{ik}B_{il} \big)\\
&-  \log B_{ii} + \lambda \sum_{k = 1, k \neq i}^{m_{r}} \abs{B_{ik}} \Big) 
\end{align*}
only depends on the terms in block row $r$. As a result, the minimization of each block row can be implemented in parallel as shown in Algorithm \ref{alg:pdagR}. This can lead to huge computational advantages, as illustrated in our experiments. 
\item (Number of computations) With the additional partition information, many of the entries in $B$ are automatically set to zero. This reduces the number of computations Partition-DAG needs to do in comparison to CCDr. In addition, many of the computations we carry out for Partition-DAG fall under Case II as discussed in Section \ref{sec:two:blocks}, which is much simpler and faster than computations under Case III, which is what CCDr needs to do for every single coordinate. 
\item (Estimation Accuracy) This is very obvious, but leveraging more information can lead to a improved estimation accuracy as borne out in our simulations studies. 
\end{enumerate}


\section{Simulation experiments} \label{simulations}

\noindent
In this section, we perform extensive simulations using four algorithms: CCDr (no ordering information is assumed to be 
known), PC algorithm (see \cite{Kalisch:Buhlmann:2007}, no ordering information is assumed to be known), the 
proposed Partition-DAG (partial ordering information is assumed to be known), and CSCS (assumes full ordering is known). 
The goal is to understand/explore the following questions about the Partition-DAG algorithm in realistic settings. 
\begin{itemize}
\item Can Paritition-DAG effectively leverage the partial ordering information to improve performance (as compared to 
methods such as CCDr and PC algorithm which do not use any ordering information)?  
\item As the partitions become finer, does the performance of Partition-DAG improve?
\item If the number of sets in the partition is kept the same, but the elements of these sets are changes so that the 
partition is more informative, then does the performance of Partition-DAG improve? 
\item How does the computational speed of Partition-DAG compare to CCDr, and how does this change when the 
partition becomes finer? 
\end{itemize}

\noindent
We investigate each of these questions separately in the subsections below. The testing based PC algorithm is much 
slower than the penalized algorithms (CSCS, Partition-DAG, CCDr), but serves as a popular benchmark, and hence is 
included in the first two subsections below when the number of network nodes is $50$. In Section \ref{randomdags}, 
we consider networks with $100$ and $200$ nodes, and the PC algorithm can be very slow in these settings. Hence, 
we do not include it in Section \ref{randomdags}.

\subsection{Partial ordering info vs. no ordering info: DREAM3 data} \label{partial:ordering:no:ordering}

\noindent
The goal of this experiment is to explore if partition based ordering information can help improve accuracy and computational 
efficiency in realistic settings. With this in mind, we perform a number of simulation studies using gene regulatory networks 
from the DREAM3 In Silico challenge \cite{dream1, dream2, dream3}. This challenge provides the transcriptional networks 
for three  yeast organisms, which we will denote as Yeast 1, Yeast 2, Yeast 3. These networks mimic activations and 
regulations that occur in gene regulatory networks. All networks are known and have 50 nodes. 

For each DAG, we generated a random $B$ by sampling the off-diagonal non-zero terms from a uniform distribution 
between 0.3 and 0.7 and assigned them a positive or negative sign with equal probability. The diagonal terms were all set to 
1. Then, the ``true" $\Omega = B^t B$ was computed, and the corresponding multivariate Gaussian distribution was used to 
generate twenty datasets each for sample size $n \in \{40,50,100,200\}$. For each sample size, each of the three 
algorithms (PC, CCDr, Partition-DAG) was run for each dataset described above for a range of penalty parameter values 
(for the PC algorithm, the significance level for the hypothesis tests was used as a penalty parameter). Note that the DAG 
estimation problem is a three way classification problem (two classes corresponding to two kinds of directed edges, and 
one class corresponding to no edge). Hence, the performance was summarized using the corresponding mean AUC-MA 
(Macro-averaged Area-Under-the-Curve, see for example \cite{TKV:2010}) value over the twenty repetitions. For each 
method, the appropriate penalty parameter (or significance level in the case of the PC algorithm) was varied over a range 
of 30 values to yield a completely sparse network at one end, and a completely dense network (ignoring edge directions) 
at the other end. The AUC values for the three binary classification problems (corresponding to each class) were then 
computed and normalized by dividing with the respective false positive rate (FPR) range. The AUC-MA was then computed 
by taking the average of the three class-wise AUC values. The results for the three different networks (Yeast1, Yeast2, 
Yeast3) are summarized in Table \ref{yeastauc}. 

As expected, the performance of each method improves with increasing sample size. It is clear that Partition-DAG 
outperforms both the PC algorithm and the CCDr algorithm for all the DAGs by quite a large margin and the performance
is more or less consistent regardless of the complexity of the DAG topology. This demonstrates that the Partition-DAG 
algorithm can successfully leverage the partial ordering information to improve performance. 
\begin{table}[!h]
\centering
\begin{tabular}{lrrrrr|}
\hline
Method & Sample size  & AUC-MA (Yeast 1) & AUC-MA (Yeast 2) & AUC-MA (Yeast 3)\\ 
\hline
CCDr & 40 & 0.5097 & 0.4413 & 0.4397\\
Partition-DAG & 40 & 0.6470 & 0.5842 & 0.5619\\ 
PC & 40 & 0.4743& 0.4160 & 0.4211\\
\hline
CCDr & 50 & 0.5242 & 0.4505 & 0.4489\\
Partition-DAG & 50 & 0.6548 & 0.6108 & 0.5767\\
PC & 50 & 0.5009 & 0.4273 & 0.4278\\
\hline
CCDr & 100 & 0.5624 & 0.4815 & 0.4716\\
Partition-DAG & 100 & 0.6830 & 0.6302 & 0.6004\\
PC & 100 & 0.5470 & 0.4491 & 0.4576\\
\hline
CCDr & 200 & 0.5884 & 0.4943 & 0.4895\\
Partition-DAG & 200 & 0.7070 & 0.6544 & 0.6150\\
PC & 200 & 0.5795 & 0.4780 & 0.4822\\
\hline
\end{tabular}
\caption{Macro Averaged Area-Under-the-Curve (AUC-MA) values for Yeast 1, Yeast 2, and Yeast 3 networks 
for CCDr, Partition-DAG and PC algorithm}
\label{yeastauc}
\end{table}

\subsection{Fine partition vs. coarse partition: Yeast 3 network} \label{fine:partition:coarse:partition}

\noindent
Note that given a partition, Partition-DAG assumes the ordering of edges between the sets 
in the partition to be known, and the ordering of edges within each partition set to be 
unknown). The goal of this subsection is to explore if using a finer partition (and hence 
more knowledge of the ordering) improves the performance of the Partition-DAG approach. 
We perform simulations using the Yeast 3 data from the DREAM3 challenge 
mentioned in Section \ref{partial:ordering:no:ordering}. Recall that the underlying 
network is known, mimics activations and regulations that occur in gene regulatory 
networks and has $p = 50$ nodes. We topologically order the nodes from $1$ to $50$, 
so that any edge directs from a smaller node to a bigger node. Again, we construct a 
``true" $\Omega$ matrix consistent with the known DAG, and generated hundred 
multivariate datasets each for sample size $n \in \{40, 50, 100, 200\}$. We analyze each of the $400$ 
datasets thus generated using four methods: Partition-DAG with a partition consisting 
of two sets: one with nodes $1$ to $36$, and the other one with nodes $37$ to $50$ 
(refered to as PDAG-2), Partition-DAG with a 
partition consisting of three sets: one with the first $24$ nodes, one with nodes $25$ to $36$, 
and one with nodes $37$ to $50$ (refered to as PDAG-3), Partition-DAG with a partition 
consisting of four sets with one with the first $12$ nodes, one with nodes $13$ to $24$, one with 
nodes $25$ to $36$, and one with nodes $37$ to $50$. The Macro-averaged Area-Under-the-Curve 
(AUC-MA) values for each algorithm are provided in Table \ref{yeastdag}. 

Again, we see that as expected, the performance of each method improves with 
increasing sample size. Also, as more information about the ordering becomes 
available with finer partitions, the performance of Partition-DAG clearly improves. 
\begin{table}[!h]
\centering
\begin{tabular}{lrrr|}
\hline
Method & Sample size & AUC-MA\\ 
\hline
PC & 40 & 0.4163\\ 
CCDR & 40 & 0.4381\\ 
PDAG-2 & 40 & 0.5714\\ 
PDAG-3 & 40 & 0.5740\\ 
PDAG-4 & 40 & 0.5970\\
\hline
PC & 50 & 0.4247\\ 
CCDR & 50 & 0.4420\\ 
PDAG-2 & 50 & 0.5792\\ 
PDAG-3 & 50 & 0.5970\\ 
PDAG-4 & 50 & 0.6189\\
\hline
PC & 100 & 0.4540\\ 
CCDR & 100 & 0.4597\\ 
PDAG-2 & 100 & 0.5987\\ 
PDAG-3 & 100 & 0.6069\\ 
PDAG-4 & 100 & 0.6393\\
\hline
PC & 200 & 0.4756\\ 
CCDR & 200 & 0.4757\\ 
PDAG-2 & 200 & 0.6112\\ 
PDAG-3 & 200 & 0.6277\\ 
PDAG-4 & 200 & 0.6585\\ 
\hline
\end{tabular}
\caption{Macro Averaged Area-Under-the-Curve (AUC-MA) values for the Yeast 3 network for Partition-DAG with various 
partition sizes (PDAG-$j$ refers to a partition consisting of $j$ sets, PC refers to the PC algorithm)}
\label{yeastdag}
\end{table}

\subsection{Informative vs. Non-informative partition: Yeast 1 network} 
\label{informative:ordering:noninformative:ordering}

\noindent
In this section, we compare the performance of Partition-DAG using two different partitions for the 
Yeast 1 network. Recall that this network has $50$ nodes. We topologically order the nodes 
from  $1$ to $50$ so that any edge in the true network directs from a smaller node to a larger 
node. The first partition consists of two sets: $V_1 = \{4, 9, 11, 16, 24\}$ and its complement. The 
second partition also consists of two sets: $V_2 = \{9, 10, 11, 13, 16\}$ and its complement. 
The partitions are constructed such that in the true graph, any edge between $V_i$ and $V_i^c$ 
directs from $V_i$ to $V_i^c$ for $i = 1,2$. The first partition is more ``informative" in the sense 
that in the true network, more edges exist between $V_1$ and $V_1^c$ as compared to edges 
between $V_2$ and $V_2^c$. Similar to earlier subsections we generated hundred 
multivariate datasets (from a multivariate Gaussian consistent with the true network) each for 
sample size $n \in \{50, 100, 200\}$, and applied Partition-DAG with the two different 
partitions discussed above (referred to as PDAG-INFO and PDAG-NONINFO) for a range of 
penalty parameter values. The Macro-averaged Area-Under-the-Curve (AUC-MA) values are 
provided in Table \ref{table:info:noninfo}. These results show that the performance of 
Partition-DAG improves as we go from an non-informative ordering to an informative ordering, 
but the difference between the two AUC values grows smaller with increasing sample size. 
\begin{table}[!h]
\centering
\begin{tabular}{rrr}
\hline
Method & Sample size & AUC-MA\\ 
\hline
PDAG-INFO & 40 & 0.6272\\ 
PDAG-NONINFO & 40 & 0.5864\\ 
\hline
PDAG-INFO & 50 & 0.6321\\ 
PDAG-NONINFO & 50 & 0.6025\\ 
\hline
PDAG-INFO & 100 & 0.6617\\ 
PDAG-NONINFO & 100 & 0.6263\\
\hline
PDAG-INFO & 200 & 0.6801\\
PDAG-NONINFO & 200 & 0.6401\\
\hline
\end{tabular}
\caption{Macro Averaged Area-Under-the-Curve (AUC-MA) values for the Yeast 1 network for Partition DAG with informative (PDAG-INFO) vs. 
non-informative (PDAG-NONINFO) partitions}
\label{table:info:noninfo}
\end{table}

\subsection{Scaling the number  of nodes and computational time} \label{randomdags}

\noindent
In this section, we consider networks with higher number of nodes ($p = 100$ and $p = 200$), and demonstrate that 
the parallelizability for Partition-DAG helps improve computational speed as well as statistical accuracy as the 
partitions grow finer. To this end, we generated a ``true" $B$ matrix of size $100 \times 100$ with a random sparsity pattern 
of about 95\% following DAG restrictions and conditions similar to those mentioned in previous subsections. We then set 
$\Omega = B^t B$ and generated 20 datasets with multivariate normal distribution with 
mean 0 and variance $\Sigma = \Omega^{-1}$. The  nodes were topologically ordered and partitioned into 4 sets of 
equal size: $V_1 \to V_2 \to V_3 \to V_4$. For each dataset, the following algorithms were used: Partition-DAG with two 
sets $V_1$ and $V_2 \cup V_3 \cup V_4$ (referred to as PDAG-2), Partition-DAG with three sets $V_1, V_2$ and $V_3 
\cup V_4$ (referred to as PDAG-3), Partition-DAG with four sets $V_1, V_2, V_3$ and $V_4$ (refered to as PDAG-4) and 
the PC algorithm. Table \ref{randomtime} provides the average wall-clock time needed for each algorithm. It is clear that the 
time improves drastically as we consider finer partitions, which is partly due to the fact that we are doing more of the 
processing in parallel for finer partitions. Table \ref{randomarea} provides the AUC-MA values, and as expected PDAG-4 
performs the best, followed by PDAG-3 and then PDAG-2. 
\begin{table}[!h]
\centering
\begin{tabular}{rrrrrrr}
\hline
p & n & PDAG-2 & PDAG-3 & PDAG-4 & PC\\
\hline
100 & 50 & 4.5569 & 2.4027 & 0.8683 & 13.6724\\ 
100 & 75 & 4.1270 & 2.1949 & 0.7903 & 23.1989\\
100 & 100 & 3.601 & 1.8453 & 0.7246 & 29.4618\\ 
100 &  200 & 3.1554 & 1.7055 & 0.6788 & 65.4514\\
\hline
\hline
200 & 100 & 112.2623 & 57.6974 & 21.2002& 110.9797\\ 
200 & 150 & 117.5930& 57.9034 & 19.3711 & 131.0355\\
200 & 200 & 109.9690 & 53.0327 & 19.6453 & 150.8217\\ 
200 &  400 & 98.9042 & 51.0516 & 20.1499 & 186.0643\\
\hline
\end{tabular}
\caption{Average wall clock time for randomly generated DAGs with $100$ nodes for Partition-DAG with partition containing 
two sets (PDAG-2), Partition-DAG with partition containing three sets (PDAG-3), Partition-DAG with partition containing four 
sets (PDAG-4), and the PC algorithm}
\label{randomtime}
\end{table}

\begin{table}[!h]
\centering
\begin{tabular}{rrrrrrr}
\hline
p & n & PDAG-2 & PDAG-3 & PDAG-4 & PC\\
\hline
100 & 50 & 0.6777 & 0.7023 & 0.7260 & 0.5107\\
100 & 75 & 0.7083 & 0.7365 & 0.7586 & 0.5596\\ 
100 & 100 & 0.7231& 0.7582 & 0.7799 & 0.5921\\
100 &  200 & 0.7462 & 0.7686 & 0.7906 & 0.6565\\
\hline
\hline
200 & 100 & 0.6870 & 0.7167 &0.7323 & 0.4668\\
200 & 150 & 0.7058 & 0.7382 & 0.7567 & 0.4917\\ 
200 & 200 & 0.7168 & 0.7475 & 0.7646 & 0.5068\\
200 &  400 & 0.7364 & 0.7660 & 0.7840 & 0.5352\\
\hline
\end{tabular}
\caption{Macro Averaged Area-Under-the-Curve (AUC-MA) for randomly generated DAGs with $100$ and $200$ nodes for 
Partition-DAG with partition containing two sets (PDAG-2), Partition-DAG with partition containing two sets (PDAG-3), 
Partition-DAG with partition containing four sets (PDAG-4), and the PC algorithm}
\label{randomarea}
\end{table}

\section{Analysis of dairy cattle data} \label{dairydataunoc}

\subsection{Data background}

\noindent
In a recent research project led by the Universidad Nacional de Colombia and aimed to increase productivity levels in high-tropic dairy operations, 
data on $36$ economic and biological variables associated with dairy cattle operations were 
collected from $375$ high-tropic dairy farms in the municipality of Guatavita, Department of Cundinamarca, Colombia spanning the period from
June, 2016 to August, 2017. A list of the variables, along with associated acronyms is provided in Table \ref{agro:variable:list} 
. Based on domain-specific knowledge, this set of variables can be split into several groups according to 
causal relationships, such that the causal relationship between variables from different groups is known, but between 
variables from the same group is unknown. Therefore, Partition-DAG is really well-suited for analyzing this dataset, as it 
allows us to effectively incorporate this information. 
\begin{table}[!h]
\centering
\begin{tabular}{|lr|r||}
\hline
SYMBOL & DETAILED MEANING\\
\hline
``PGR" & Pasture growth rate\\
``AP" & Total pasture (hectare) area of each herd\\
``SR" &  Stocking rate, i.e., no of individuals per hectare\\
``OF" &  Amount of offered forage (kg per individual)\\
``AFI" &  Average forage intake (kg per individual)\\
``TFI" &  Average total (forage + supp) feed intake (kg per individual)\\
``AMC" &  Average of milking cows in the herd\\
``TCH" &  Total number of cows in the herd\\
``AMY" &  Average milk yield (lt) per cow per day\\
``ATS" &  Average total solids of milk (\%)\\ 
``NW" & Number of workers in the herd per month\\
``SM" &  Amount of sold milk (lt) per month\\
``PM" &   Amount of milk (lt) produced per month\\ 
``TMS" &  Total solids of milk (kg) produced per month\\  
``MH" &  Amount of milk per hectare per month\\
``TS" &  Total solids produced per hectare per day\\ 
``TSC" &  Total solids produced per cow per month\\
``AMW" &  Amount of milk (lt) per worker per month\\
``TSW" &  Total solids of milk (kg) per worker per month\\
``CSC" &  Cost (Colombian pesos) of soil correction strategies\\
``PMC" &  Nutritional and pasture management cost (Colombian pesos)\\ 
``RMC" &  Reproductive management cost (Colombian pesos)\\
``SMC" &  Sanitary management cost (Colombian pesos)\\
``BCC" &  Investments and bank credits cost (Colombian pesos)\\
``TAX" &  Taxes (Colombian pesos)\\
``EVM" &  Economic value of milk (Colombian pesos per litter)\\
``TC" &  Total cost\\  
``MSI" &  Total income from milk sellings (Colombian pesos)\\
``AI" &   Additional income (Colombian pesos)\\
``MPC" &  Milk production cost per litter\\
``TI" &  Total income (Colombian pesos)\\
``WC" &  Cost per worker per month (Colombian pesos)\\
``CC" &  Cost per cow per month (Colombian pesos)\\  
``CH" &  Cost per hectare per month (Colombian pesos)\\  
``NI" &   Net income (Colombian pesos)\\
``WI" & Income per worker per month (Colombian pesos)\\
\hline
\end{tabular}
\caption{Detailed list of dairy agroecosystem variables used in the analysis}
\label{agro:variable:list}
\end{table}

\subsection{Results and discussion}

\noindent
The variables were divided into 10 groups on the basis of domain-specific knowledge by an expert in dairy science 
(Dr. Carulla, co-author). Specifically, knowledge on the hierarchical structure of a grazing dairy cattle operation was 
used \cite{EDT:2006}. Colombian dairy operations are based on grasslands; thus, animals are mostly fed fresh 
forages. This hierarchy follows the natural production flow, it all starts at the soil and environmental (temperature, light, rain) 
levels, which highly affect the amount and quality of forage \cite{Dillon:2006}. Then, several variables associated to pasture 
management, cows' supplementation, and their genetic makeup determine efficiency in the complex process of transforming 
forage into milk \cite{BMKD:2003}. Since an animal production system is a business, there is always interest in maximizing 
profit, which is explained by a large number of variables and their interaction, but is summarized by a simple number: net 
income. Therefore, at the end of the hierarchy, that is, in the final causal groups, we find economic variables, basically, costs 
and incomes. 

Consequently, pasture growth rate was in the first causal group because this variable can be thought of as an output of the 
interaction between soil, environment, pasture genetics and pasture management and it highly determines the stocking rate 
(number of individuals or live weight per unit of area). Along with pasture growth rate, total grazing area defines the number 
of cows a herd can hold; as a result, the first group comprised of these two variables. Stocking rate is computed as the 
total number of individuals or total live weight in the herd divided by total grazing area; hence, stocking rate was assigned to the 
second group. The following groups contain variables associated to forage allowance and total feed intake, milk yield, and 
resources used in milk production such as number of workers. The last group comprises of two relevant economic 
variables: total net income and net income per worker. 

\begin{figure}[H]
\centering
\includegraphics[height=6in]{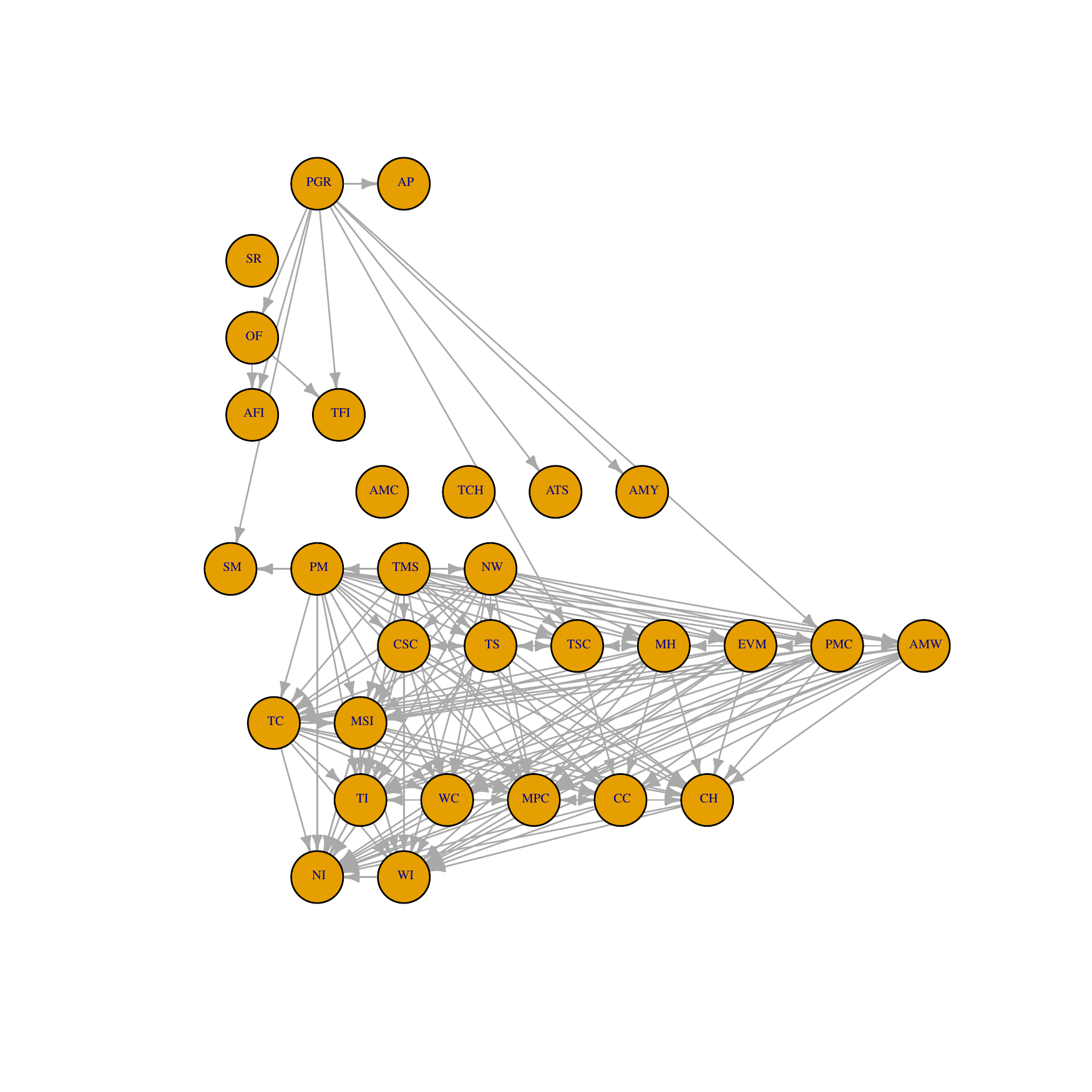}
\caption{Estimated network for dairy cattle variables using partition with 10 groups}
\label{dairy10}
\end{figure}

The penalty parameter for the Partition-DAG approach was chosen to obtain an approximate edge density around 
$0.33$ (this density requirement was defined using preliminary analysis). The estimated network (DAG) is depicted
in Figure \ref{dairy10}. 

\begin{figure}[H]
\centering
\includegraphics[height=6in]{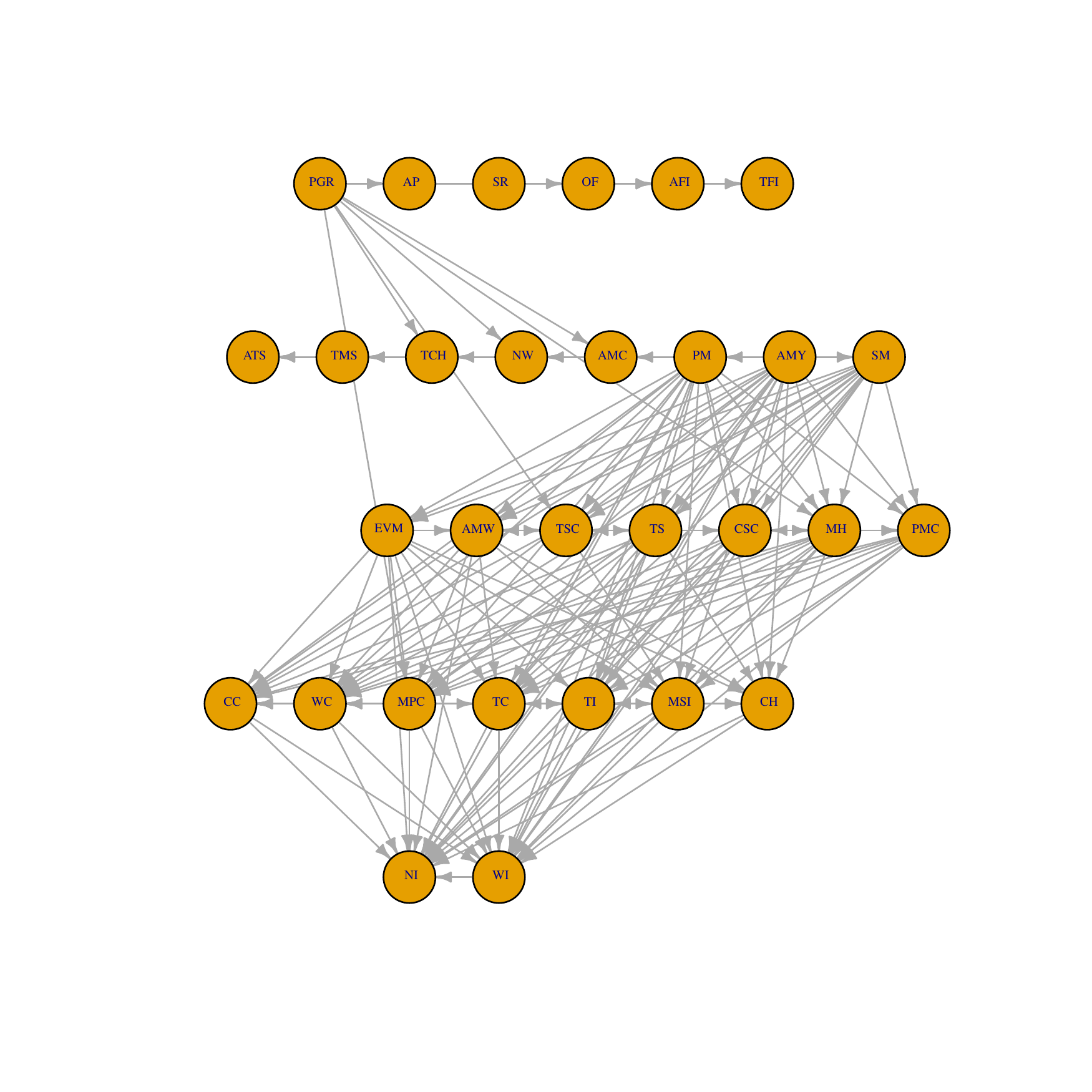}
\caption{Estimated network for dairy cattle variables using partition with 5 groups}
\label{dairy5}
\end{figure}

In order to understand/illustrate the difference in the performance of Partition-DAG with changes in the allocation of variables to partitions,
we merged some of the $10$ groups to obtain a partition with $5$ groups, and again estimated a causal network 
using Partition-DAG. The estimated network with five groups is depicted in Figure \ref{dairy5}. 

In addition to this partial ordering information, some relationships were known to exist; specifically, $38$ edges in total are known 
based on background information dictated by animal sciences theory, but also because 
some variables were explicitly computed using others. For the sake of clarity, we provide selected example next.
It is well known that total dry matter intake ($TFI$) has a direct impact on milk yield ($AMY$) \cite{HPS:2004}; in 
addition, the sign of this relationship is known to be positive. This last fact takes importance when evaluating the 
corresponding entries of the estimated Cholesky factor of the precision matrix (given their regression interpretation). 
Moreover, for economic variables it is easier to know what edges should exist in the estimated DAG because many 
economic indices are computed as linear combinations of other variables in the system. For instance, total production cost 
($TC$) is the sum of partial costs such as reproductive management cost ($RMC$) 
and cost of soil correction strategies ($CSC$). Another example is net income 
($NI$) which is the difference between total income ($TI$) and total cost 
($TC$). 

Since this extra information was not used by Partition-DAG, it allows us to carry out a {\it partial evaluation} of the 
performance of the proposed method. To this end, for the $5$ group case, $14$ out of the $31$ edges ($45\%$) were present in 
the estimated network, while in the $10$ group case, $17$ out of the $31$ edges ($55\%$) were present in the 
estimated network. The specific list of these edges and those present in the two estimated networks 
is given in Table \ref{dairy:known:edge}. 
\begin{table}[!h]
\centering
\begin{tabular}{|r|r|r|}
\hline
Edge ($a \rightarrow b$) & Five Groups network & Ten Groups	network\\	
\hline
( NW , AMW ) & Pres & Pres \\ 
  ( PM , AMW ) & Abs & Pres \\ 
  ( TFI , AMY ) & Abs & Abs \\ 
  ( TC , CC ) & Abs & Pres \\ 
  ( TCH , CC ) & Abs & Abs \\ 
  ( AP , CH ) & Abs & Abs \\ 
  ( TC , CH ) & Pres & Pres \\ 
  ( NW , CSC ) & Pres & Pres \\ 
  ( TMS , CSC ) & Pres & Abs \\ 
  ( AP , MH ) & Abs & Abs \\ 
  ( PM , MH ) & Abs & Pres \\ 
  ( SR , MH ) & Abs & Abs \\ 
  ( PM , MPC ) & Abs & Pres \\ 
  ( TC , MPC ) & Abs & Pres \\ 
  ( EVM , MSI ) & Pres & Pres \\ 
  ( TC , NI ) & Pres & Pres \\ 
  ( TI , NI ) & Pres & Pres \\ 
  ( SR , OF ) & Pres & Abs \\ 
  ( AMY , PM ) & Abs & Abs \\ 
  ( PMC , TC ) & Abs & Pres \\ 
  ( SR , TCH ) & Abs & Abs \\ 
  ( AFI , TFI ) & Abs & Abs \\ 
  ( MSI , TI ) & Pres & Pres \\ 
  ( ATS , TMS ) & Abs & Abs \\ 
  ( AP , TS ) & Abs & Abs \\ 
  ( TMS , TS ) & Pres & Abs \\ 
  ( TMS , TSC ) & Pres & Abs \\ 
  ( NW , WC ) & Pres & Pres \\ 
  ( TC , WC ) & Abs & Pres \\ 
  ( NW , WI ) & Pres & Pres \\ 
  ( TI , WI ) & Pres & Pres \\ 
\hline
\end{tabular}
\caption{List of more known edges between dairy agroecosystem variables (in addition to the partition information), and 
whether these edges were present in the estimated networks}
\label{dairy:known:edge}
\end{table}

Also, a dairy science expert (Dr. Carulla, co-author) inspected all the estimated edges for each network, and identified which 
of the estimates edges were not meaningful. The following are some examples of incorrectly inferred edges because a direct 
causal relationship between the corresponding pair of variables does not exist: the economic value of milk ($EVM$) 
and production cost per liter ($MPC$), pasture growth rate (Growth rate) and total pasture area 
($AP$), and growth rate and number of workers per month ($NW$). As to the first edge, total 
milk cost per litter is determined by total production cost and total milk yield, not by the economic value of milk. The price 
paid to the producer does not cause production cost. Regarding the second edge, it is not reasonable to state that total 
grazing area is caused by pasture growth rate. Total grazing area is explained by herd size and its productive targets (e.g., 
dairy and other agricultural activities), not by how fast its pastures grow. Finally, pasture growth rate cannot have a direct 
effect on the number of workers per month, because the later variable depends on social factors as well as on the 
productivity level, of course, growth rate affects ?productivity level and therefore, it could have an impact on the number of 
workers, but this does not imply a direct causal relationship. To summarize, for the $5$ group network, $19\%$ of the 
estimated edges were not meaningful, while for the $10$ group network, $17\%$ of the estimated edges were not 
meaningful. Once again, it can be seen that Partition-DAG is able to leverage additional external information (in the form of a 
finer partition) to improve performance. 

The above analysis illustrates the tremendous usefulness of the application of the proposed method in dairy science and in 
the problem of DAG estimation more generally. In the first place, it is a sound approach to infer a DAG and the associated causal relationships
from observational data in 
animal production systems using the available (partial) information, which is a feature exhibited by most of 
these agroecosystems. Further, knowledge of causal relationships in dairy and other animal production 
operations helps in taking management decisions and making recommendatios, since it leads to identifying what nodes 
have the biggest impact on the system outputs; the latter information singles out variables that should be modified and the direction of 
such modifications. Finally, the analysis of the estimated DAGs sheds light into interesting 
interactions between components of the production system and thus could be the basis to design follow-up experiments to fully test their validity.

\bibliographystyle{plainnat}
\bibliography{references}

\begin{thebibliography}{34}
\providecommand{\natexlab}[1]{#1}
\providecommand{\url}[1]{\texttt{#1}}
\expandafter\ifx\csname urlstyle\endcsname\relax
  \providecommand{\doi}[1]{doi: #1}\else
  \providecommand{\doi}{doi: \begingroup \urlstyle{rm}\Url}\fi

\bibitem[Altamore et~al.(2013)Altamore, Consonni, and
  Rocca]{Altamore:Consonni:2013}
D.~Altamore, G.~Consonni, and L.~La Rocca.
\newblock Objective {B}ayesian search of {G}aussian directed acyclic graphical
  models for ordered variables with non-local priors.
\newblock \emph{Biometrics}, 69:\penalty0 478--487, 2013.

\bibitem[Aragam and Zhou(2015)]{Aragam:Zhou:2015}
B.~Aragam and Q.~Zhou.
\newblock Concave penalized estimation of sparse {G}aussian {B}ayesian
  networks.
\newblock \emph{Journal of Machine Learning Research}, 16:\penalty0 2273--2328,
  2015.

\bibitem[Bargo et~al.(2003)Bargo, Muller, Kolver, and Delahoy]{BMKD:2003}
F.~Bargo, L.~D. Muller, E.~S. Kolver, and J.~E. Delahoy.
\newblock Invited review: production and digestion of supplemented dairy cows
  on pasture.
\newblock \emph{J Dairy Sci.}, 86:\penalty0 1--42, 2003.

\bibitem[Cao et~al.(2017)Cao, Khare, and Ghosh]{CKG:2017}
Xuan Cao, Kshitij Khare, and Malay Ghosh.
\newblock Posterior graph selection and estimation consistency for
  high-dimensional {B}ayesian {D}{A}{G} models.
\newblock \emph{https://arxiv.org/abs/1611.01205}, 2017.

\bibitem[Chickering(2003)]{C:2003}
David~Maxwell Chickering.
\newblock Optimal structure identification with greedy search.
\newblock \emph{The Journal of Machine Learning Research}, 3:\penalty0
  507--554, 2003.

\bibitem[Consonni et~al.(2017)Consonni, Rocca, and Peluso]{Consonni:2017}
G.~Consonni, L.~La Rocca, and S.~Peluso.
\newblock Objective {B}ayes covariate-adjusted sparse graphical model
  selection.
\newblock \emph{Scand. J. Statist}, 44:\penalty0 741--764, 2017.

\bibitem[Dillon(2006)]{Dillon:2006}
P.~Dillon.
\newblock Achiving high dry-matter intake from pastures with grazing dairy
  cows.
\newblock In A.~Elgersma, J.~Dijkstra, and S.~Tamminga, editors, \emph{J.
  Dijkstra and S. Tamminga Fresh Herbage for Dairy Cattle, 1-26. Springer.
  Printed in the Netherlands}. Springer, Netherlands, 2006.

\bibitem[Elgersma et~al.(2006)Elgersma, Dijkstra, and Tamminga]{EDT:2006}
A.~Elgersma, J.~Dijkstra, and S.~Tamminga.
\newblock \emph{Fresh Herbage for Dairy Cattle}.
\newblock Springer, Netherlands, 2006.

\bibitem[Ellis and Wong(2008)]{EW:2008}
Byron Ellis and Wing~Hung Wong.
\newblock Learning causal {B}ayesian network structures from experimental data.
\newblock \emph{Journal of the American Statistical Association}, 103\penalty0
  (482):\penalty0 778--789, 2008.

\bibitem[Emmert-Streib et~al.(2014)Emmert-Streib, Dehmer, and
  Haibe-Kains]{emmert2014gene}
Frank Emmert-Streib, Matthias Dehmer, and Benjamin Haibe-Kains.
\newblock Gene regulatory networks and their applications: understanding
  biological and medical problems in terms of networks.
\newblock \emph{Frontiers in Cell and Developmental Biology}, 2:\penalty0 38,
  2014.

\bibitem[Gamez et~al.(2011)Gamez, Mateo, and Puerta]{G:2011}
Jose~A Gamez, Juan~L Mateo, and Jose~M Puerta.
\newblock Learning bayesian {N}etworks by hill climbing: Efficient methods
  based on progressive restriction of the neighborhood.
\newblock \emph{Data Mining and Knowledge Discovery}, 22\penalty0
  (1-2):\penalty0 106--148, 2011.

\bibitem[Gamez et~al.(2012)Gamez, Mateo, and Puerta]{G:2012}
Jose~A Gamez, Juan~L Mateo, and Jose~M Puerta.
\newblock One iteration chc algorithm for learning bayesian networks: An
  effective and efficient algorithm for high dimensional problems.
\newblock \emph{Progress in Artificial Intelligence}, 1\penalty0 (4):\penalty0
  329--346, 2012.

\bibitem[Geiger and Heckerman(2013)]{GH:2013}
Dan Geiger and David Heckerman.
\newblock Learning {G}aussian networks.
\newblock \emph{arXiv:1302.6808}, 2013.

\bibitem[Heckerman et~al.(1995)Heckerman, Geiger, and Chickering]{H:1995}
David Heckerman, Dan Geiger, and David~M Chickering.
\newblock Learning {B}ayesian networks: The combination of knowledge and
  statistical data.
\newblock \emph{Machine learning}, 20\penalty0 (3):\penalty0 197--243, 1995.

\bibitem[Hristov et~al.(2004)Hristov, Price, and Shafii]{HPS:2004}
A.~N. Hristov, W.~J. Price, and B.~Shafii.
\newblock A meta-analysis examining the relationship among dietary factors, dry
  matter intake, and milk and milk protein yield in dairy cows.
\newblock \emph{J Dairy Sci.}, 87:\penalty0 2184--2196, 2004.

\bibitem[Huang et~al.(2006)Huang, Liu, Pourahmadi, and Liu]{HLPL:2006}
J.~Huang, N.~Liu, M.~Pourahmadi, and L.~Liu.
\newblock Covariance selection and estimation via penalised normal likelihoode.
\newblock \emph{Biometrika}, 93:\penalty0 85--98, 2006.

\bibitem[Jalvingh(1992)]{Jalvingh:1992}
A.W. Jalvingh.
\newblock The possible role of existing models in on-farm decision support in
  dairy cattle and swine production.
\newblock \emph{Livestock Production Science}, 31:\penalty0 355--365, 1992.

\bibitem[Kalisch and Buhlmann(2007)]{Kalisch:Buhlmann:2007}
M.~Kalisch and P.~Buhlmann.
\newblock Estimating high-dimensional directed acyclic graphs with the
  pc-algorithm.
\newblock \emph{Journal of Machine Learning Research}, 8:\penalty0 613--636,
  2007.

\bibitem[Kalisch and B¨uhlmann(2007)]{KB:2007}
Markus Kalisch and Peter B¨uhlmann.
\newblock Estimating high-dimensional directed acyclic graphs with the
  pc-algorithm.
\newblock \emph{The Journal of Machine Learning Research}, 8:\penalty0
  613--636, 2007.

\bibitem[Khare et~al.(2017)Khare, Oh, Rahman, and Rajaratnam]{KORR:2017}
Kshitij Khare, Sang Oh, Syed Rahman, and Bala Rajaratnam.
\newblock A convex framework for high-dimensional sparse {C}holesky based
  covariance estimation.
\newblock \emph{https://arxiv.org/abs/1610.02436}, 2017.

\bibitem[Lam and Bacchus(1994)]{LB:1994}
Wai Lam and Fahiem Bacchus.
\newblock Learning {B}ayesian belief networks: An approach based on the
  {M}{D}{L} principle.
\newblock \emph{Computational Intelligence}, 10\penalty0 (3):\penalty0
  269--293, 1994.

\bibitem[Marbach et~al.(2009)Marbach, Schaffter, Mattiussi, and
  Floreano]{dream3}
Daniel Marbach, Thomas Schaffter, Claudio Mattiussi, and Dario Floreano.
\newblock Generating {R}ealistic {I}n {S}ilico {G}ene {N}etworks for
  {P}erformance {A}ssessment of {R}everse {E}ngineering {M}ethods.
\newblock \emph{Journal of {C}omputational {B}iology}, 16\penalty0
  (2):\penalty0 229--239, 2009.
\newblock \doi{10.1089/cmb.2008.09TT}.
\newblock WingX.

\bibitem[Marbach et~al.(2010)Marbach, Prill, Schaffter, Mattiussi, Floreano,
  and Stolovitzky]{dream2}
Daniel Marbach, Robert~J. Prill, Thomas Schaffter, Claudio Mattiussi, Dario
  Floreano, and Gustavo Stolovitzky.
\newblock Revealing strengths and weaknesses of methods for gene network
  inference.
\newblock \emph{{PNAS}}, 107\penalty0 (14):\penalty0 6286--6291, 2010.
\newblock \doi{10.1073/pnas.0913357107}.
\newblock WingX.

\bibitem[Marbach et~al.(2012)Marbach, Costello, K{\"u}ffner, Vega, Prill,
  Camacho, Allison, Aderhold, Bonneau, Chen, et~al.]{marbach2012wisdom}
Daniel Marbach, James~C Costello, Robert K{\"u}ffner, Nicole~M Vega, Robert~J
  Prill, Diogo~M Camacho, Kyle~R Allison, Andrej Aderhold, Richard Bonneau,
  Yukun Chen, et~al.
\newblock Wisdom of crowds for robust gene network inference.
\newblock \emph{Nature methods}, 9\penalty0 (8):\penalty0 796, 2012.

\bibitem[Prill et~al.(2010)Prill, Marbach, Saez-Rodriguez, Sorger, Alexopoulos,
  Xue, Clarke, Altan-Bonnet, and Stolovitzky]{dream1}
Robert~J. Prill, Daniel Marbach, Julio Saez-Rodriguez, Peter~K. Sorger,
  Leonidas~G. Alexopoulos, Xiaowei Xue, Neil~D. Clarke, Gregoire Altan-Bonnet,
  and Gustavo Stolovitzky.
\newblock Towards a rigorous assessment of systems biology models: The
  {D}{R}{E}{A}{M}3 challenges.
\newblock \emph{PLOS ONE}, 5\penalty0 (2):\penalty0 1--18, 02 2010.
\newblock \doi{10.1371/journal.pone.0009202}.
\newblock URL \url{https://doi.org/10.1371/journal.pone.0009202}.

\bibitem[Shojaie and Michailidis(2010)]{Shajoie:Michalidis:2010}
A.~Shojaie and G.~Michailidis.
\newblock Penalized likelihood methods for estimation of sparse
  high-dimensional directed acyclic graphs.
\newblock \emph{Biometrika}, 97:\penalty0 519--538, 2010.

\bibitem[Shojaie et~al.(2014)Shojaie, Jauhiainen, Kallitsis, and
  Michailidis]{SJKM:2014}
A.~Shojaie, A.~Jauhiainen, M.~Kallitsis, and G.~Michailidis.
\newblock Inferring regulatory networks by combining perturbation screens and
  steady state gene expression profiles.
\newblock \emph{PLoS ONE}, 9:\penalty0 e82393, 2014.

\bibitem[Spirtes et~al.(2001)Spirtes, Glymour, and Scheines]{PCR:2001}
Peter Spirtes, Clark Glymour, and Richard Scheines.
\newblock \emph{Causation, Prediction, and Search}.
\newblock 2001.

\bibitem[Thornley and France(2007)]{Thornley:France:2007}
J.H.M. Thornley and J.~France.
\newblock \emph{Mathematical models in agriculture. Quantitative methods for
  Plant, Animal and Ecological sciences.}
\newblock Cromwell Press Trowbridge, UK, 2nd edition edition, 2007.

\bibitem[Tsamardinos et~al.(2006)Tsamardinos, Brown, and Aliferis]{T:2006}
Ioannis Tsamardinos, Laura~E Brown, and Constantin~F Aliferis.
\newblock The max-min hill-climbing {B}ayesian network structure learning
  algorithm.
\newblock \emph{Machine Learning, 65(1):31--78, 2006.}, 65\penalty0
  (1):\penalty0 31--78, 2006.

\bibitem[Tseng(2001)]{Tseng:2001}
P.~Tseng.
\newblock Convergence of a block coordinate descent method for
  nondifferentiable minimization.
\newblock \emph{Journal of Optimization Theory and Applications}, 109:\penalty0
  475--494, 2001.

\bibitem[Tsoumakas et~al.(2010)Tsoumakas, Katakis, and Vlahavas]{TKV:2010}
G.~Tsoumakas, I.~Katakis, and I.~P. Vlahavas.
\newblock Mining multi-label data.
\newblock In O.~Maimon and L.~Rokach, editors, \emph{Data Mining and Knowledge
  Discovery Handbook}, pages 667--685. Springer-Verlag, Heidelberg, Germany,
  2010.

\bibitem[van~de Geer and Buhlmann(2013)]{vandeGeer:Buhlmann:2013}
S.~van~de Geer and P.~Buhlmann.
\newblock l0-penalized maximum likelihood for sparse directed acyclic graphs.
\newblock \emph{Annals of Statistics}, 41:\penalty0 536--567, 2013.

\bibitem[Zhou(2011)]{Z:2011}
Qing Zhou.
\newblock Multi-domain sampling with applications to structural inference of
  {B}ayesian networks.
\newblock \emph{Journal of the American Statistical Association}, 106\penalty0
  (496):\penalty0 1317--1330, 2011.

\end{thebibliography}

\end{document}